\documentclass{article}

\usepackage{arxiv}

\usepackage[utf8]{inputenc} % allow utf-8 input
\usepackage[T1]{fontenc}    % use 8-bit T1 fonts
\usepackage{hyperref}       % hyperlinks
\usepackage{url}            % simple URL typesetting
\usepackage{booktabs}       % professional-quality tables
\usepackage{amsfonts}       % blackboard math symbols
\usepackage{nicefrac}       % compact symbols for 1/2, etc.
\usepackage{microtype}      % microtypography
\usepackage{lipsum}

\usepackage{graphicx}
\usepackage{amsmath,amssymb}
\newcommand{\mi}{\mathrm{i}}

\title{Dispersion forces in inhomogeneous planarly layered media: A one-dimensional model for effective polarisabilities}

\author{
J.~Fiedler~\thanks{Centre for Materials Science and Nanotechnology, Department of Physics, University of Oslo, P. O. Box 1048 Blindern, NO-0316 Oslo, Norway}\\
Physikalisches Institut\\Albert-Ludwigs-Universit{\"a}t Freiburg\\Hermann-Herder-Str. 3\\79104 Freiburg, Germany\\
\texttt{johannes.fiedler@physik.uni-freiburg.de}
\And
F.~Spallek\\
Physikalisches Institut\\Albert-Ludwigs-Universit{\"a}t Freiburg\\Hermann-Herder-Str. 3\\79104 Freiburg, Germany\\
\And
P.~Thiyam\\
Division of Theoretical Chemistry\\Lund University\\P.O. Box, S-22100 Lund, Sweden\\
\And
C.~Persson\\Centre for Materials Science and Nanotechnology\\Department of Physics\\University of Oslo\\P. O. Box 1048 Blindern\\NO-0316 Oslo, Norway\\
\And
M.~Bostr{\"o}m\thanks{Centre for Materials Science and Nanotechnology, Department of Physics, University of Oslo, P. O. Box 1048 Blindern, NO-0316 Oslo, Norway}\\
Department of Energy and Process Engineering\\Norwegian University of Science and Technology\\NO-7491 Trondheim, Norway\\
\And
M.~Walter\thanks{FIT Freiburg Centre for Interactive Materials and Bioinspired Technologies, Georges-K\"ohler-Allee 105, 79110 Freiburg, Germany}
\thanks{Fraunhofer IWM, W\"ohlerstrasse 11, D-79108 Freiburg i. Br., Germany}\\Physikalisches Institut\\Albert-Ludwigs-Universit{\"a}t Freiburg\\Hermann-Herder-Str. 3\\79104 Freiburg, Germany\\
\And
S.~Y.~Buhmann~\thanks{Freiburg Institute for Advanced Studies, Albert-Ludwigs-Universit{\"a}t Freiburg, Albertstr. 19, 79104 Freiburg, Germany}\\
Physikalisches Institut\\Albert-Ludwigs-Universit{\"a}t Freiburg\\Hermann-Herder-Str. 3\\79104 Freiburg, Germany
}

\begin{document}
\maketitle

\begin{abstract}
Dispersion forces such as van der Waals forces between two microscopic particles, the Casimir--Polder forces between a particle and a macroscopic object or the Casimir force between two dielectric objects are well studied in vacuum. However, in realistic situations the interacting objects are often embedded in an environmental medium. Such a solvent influences the induced dipole interaction. With the framework of macroscopic quantum electrodynamics, these interactions are mediated via an exchange of virtual photons. Via this method the impact of a homogeneous solvent medium can be expressed as local-field corrections leading to excess polarisabilities which have previously been derived for hard boundary conditions. In order to develop a more realistic description, we investigate on a one-dimensional analog system illustrating the influence of a continuous dielectric profile.  
\end{abstract}

% keywords can be removed

\section{Introduction}
Dispersion forces belong to the weakest forces in nature and are caused by the ground-state fluctuations of the electromagnetic field~\cite{McLachlan387}. 
In this description, these fluctuations induce dipole moments inside the considered objects which can interact with other materials. Alternative accounts derive dispersion forces from position dependent ground-state energies of the coupled field-matter system~\cite{Casimir48,Casimir482}.  The dispersion forces resulting from this process are classified by three different types of interacting partners: the Casimir force which describes the interaction of two neutral macroscopic dielectric objects, the van der Waals force acting between two polarisable particles and the Casimir--Polder force connecting both by describing the interaction of a polarisable particle with a macroscopic dielectric body. During recent years these forces were well studied in several experiments \cite{Grisenti1999,Arndt1999,Juffmann2012,Brand2015} and in theory \cite{Casimir48,pitaevskii,Scheel2008,Buhmann2013Dispersion}. However, all these investigations were restricted to laboratory conditions by using vacuum chambers which remove any environment. 

In natural situations, where these effects play an important role, such as the Gecko feet \cite{Gecko} or colloids \cite{Friedrich}, this assumption is not valid any more. In principle, one can assume that an environmental medium results in a decrease of the interaction caused by the screening by the medium of permittivity $\varepsilon$. This is expressed by a factor $\varepsilon^{-1}$ leading to the expected decrease of the potential.

Due to the presence of the particle, a Pauli exclusion cavity is formed \cite{messiah2014quantum} 
This has given rise to the development of a cavity model \cite{Johannes} describing the boundaries of both medium and particle as hard discontinuous changes in the permittivity. For the resulting excess polarisabilities, a spherical two- or three-layer system is considered for Onsager's 
\begin{figure}[t!]
\centering
\includegraphics[width=0.6\textwidth]{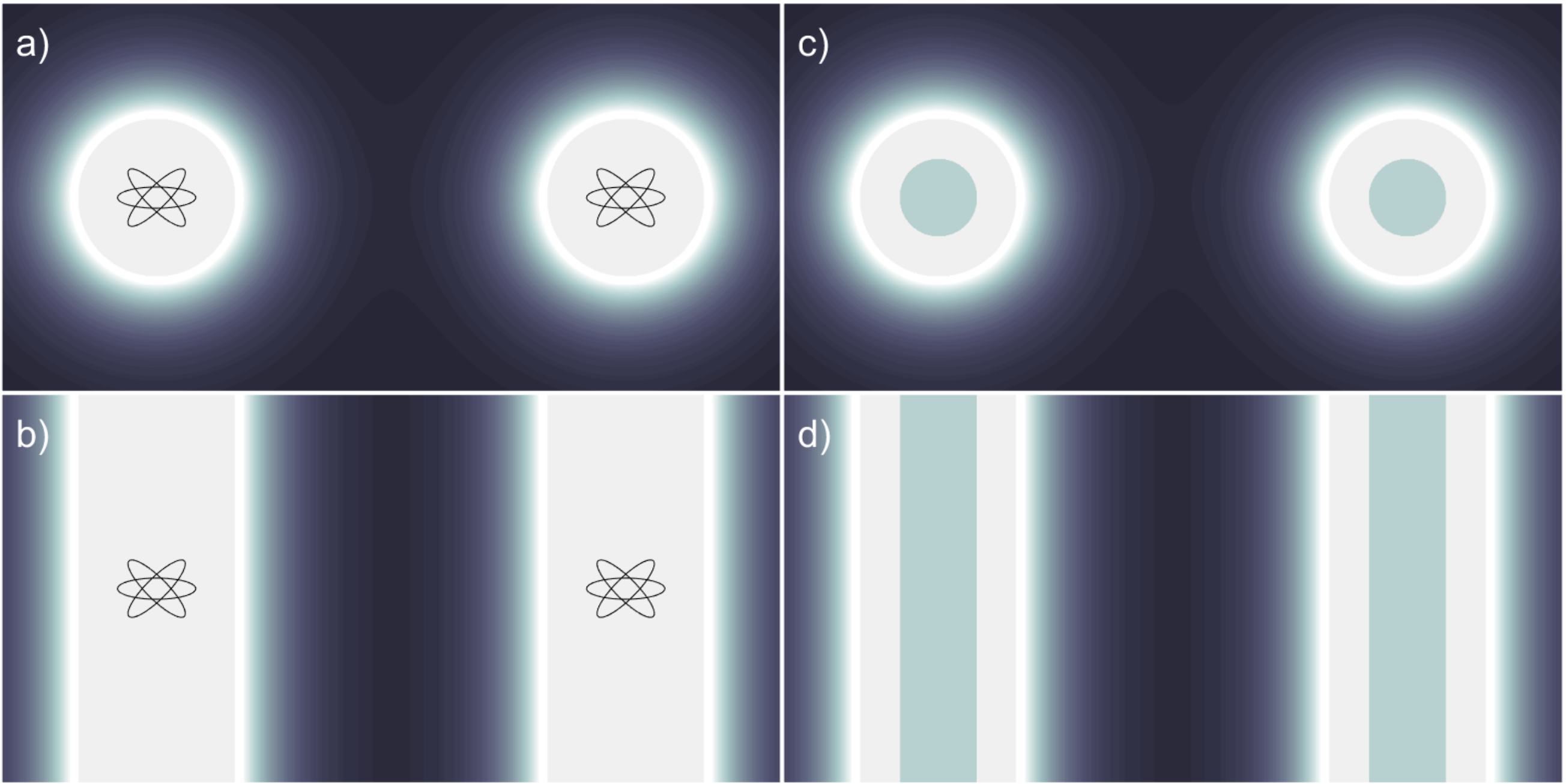}
\caption{Sketch of the considered setups for the spherical problem and the one dimensional analogon: a) Two particles embedded in a medium creating an Onsager's real cavity with inhomogeneous dielectric profile; b) one-dimensional analogon with planarly inhomogeneous profile; c) Two spherical nano-particles embedded in a medium with an inhomogeneous cavity; d) the corresponding one-dimensional problem with two dielectric plates of finite thickness embedded in a medium with inhomogeneous profile.} \label{fig:Csketch1}
\end{figure}
real cavity model and the finite-size model, respectively. In the two layer case the particle is treated to be point-like in the centre and the optical response is modified by the transmission of light through the boundary following Mie reflection (Fig.~\ref{fig:Csketch1} a). For larger objects, such as clusters or molecules, the particle's extension has to be taken into account which will be modelled by an additional layer (Fig.~\ref{fig:Csketch1} b). In this case, the excess polarisability is determined by the reflection of light at the outer boundary. 
The statistical nature of interactions between particles and
  medium (e.g. in molecules in solvents) suggests that these boundaries 
  are better described by continuous profiles~\cite{Walter14}, which were numerically investigated~\cite{Henkel2008}.

\begin{figure}[tb]
\centering
\includegraphics[width=0.6\textwidth]{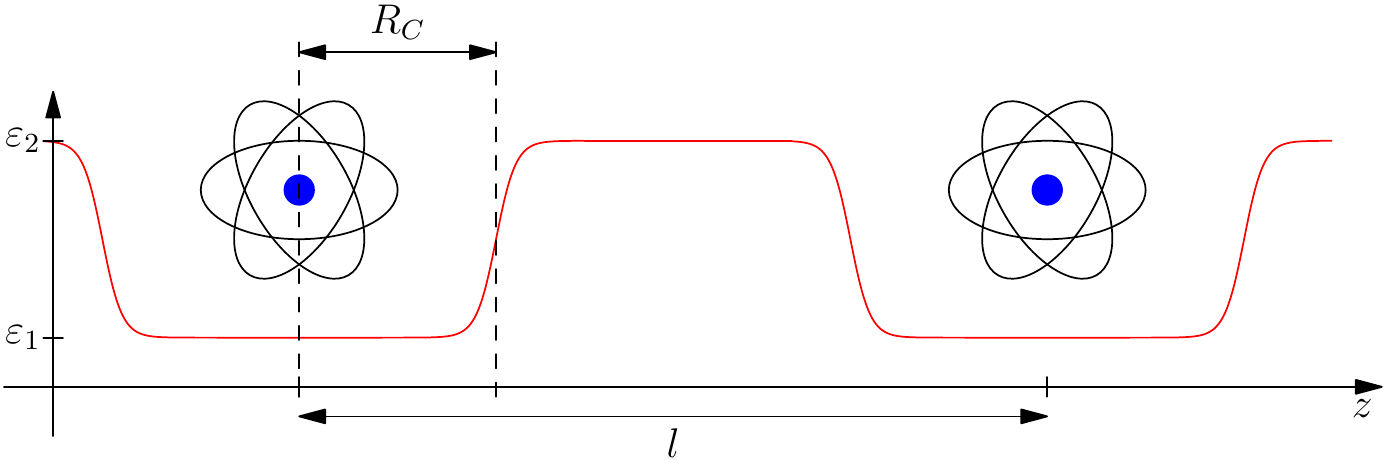}
\caption{Sketch of the arrangement for the van der Waals force between two atoms distant by $l$ and embedded in an inhomogeneous planarly layered media (red curve) with a cavity radius $R_C$.} \label{fig:Csketch2}
\end{figure}
In this manuscript, we analyse the influence of continuous boundary profiles on dispersion forces. For simplicity we consider planarly layered systems leading to an effective one-dimensional model. Figure~\ref{fig:Csketch1} illustrates the corresponding models. Two particles interacting via van der Waals forces are sketched, Fig.~\ref{fig:Csketch1}~a), which are embedded in a medium and create an inhomogeneous boundary profile. The corresponding one-dimensional situation is depicted in b), where we consider two planar cavities with continuous boundaries and centred particles interacting with each other. Figure~\ref{fig:Csketch1}~c) illustrates the finite-size cavity model with an inhomogeneous boundary. The corresponding situation consists of two plate of finite thickness embedded in a planar cavity, see Fig.~\ref{fig:Csketch1}~d).

We hence consider two scenarios: two particles or two plates in a five-layer system, representing the van der Waals and the Casimir force, respectively. In these two cases, we cover all important wave propagations at an inhomogeneous profile - the transmission through and the reflection at a continuous profile.

\section{Casimir force and van der Waals potential}
\begin{figure}[htb]
\centering
\includegraphics[width=0.6\textwidth]{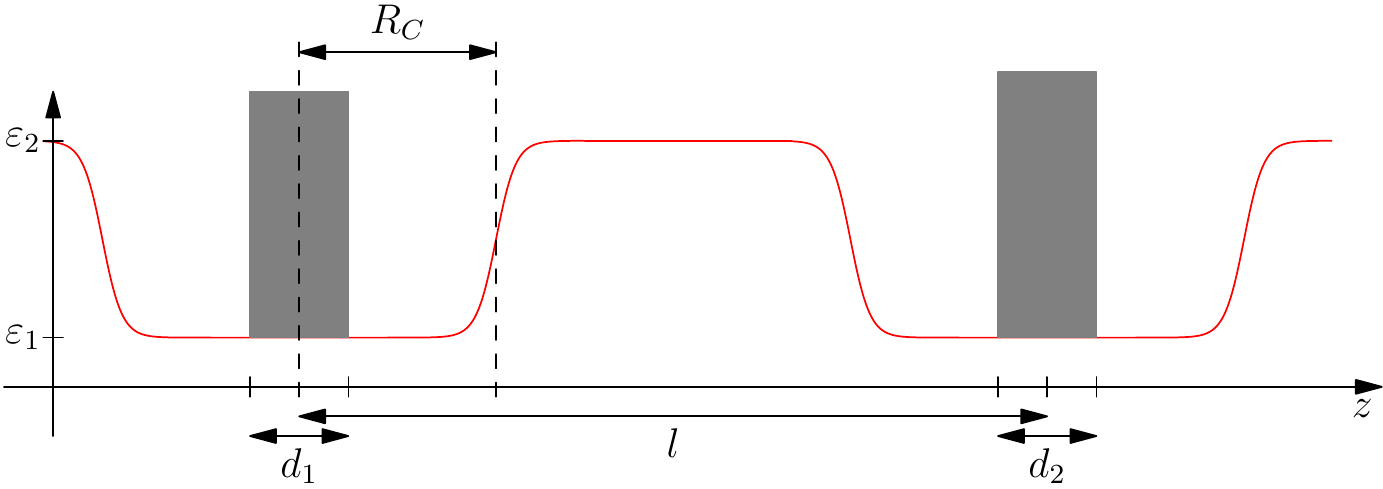}
\caption{Sketch of the arrangement for the Casimir force between two parallel plates (grey) of thickness $d_1$ and $d_2$ distant by $l$ and embedded in an inhomogeneous planarly layered media (red curve) with a cavity radius $R_C$.} \label{fig:Csketch}
\end{figure}

Figures~\ref{fig:Csketch2} and \ref{fig:Csketch} sketch a cross section along the perpendicular direction to the layers for the van der Waals and Casimir cases, respectively. Two objects are embedded in a medium of permittivity $\varepsilon_2$ and create a cavity with a vacuum permittivity $\varepsilon_1=1$. We focus on the impact of the continuous boundary conditions onto the dispersion forces. To this end, we repeat the important steps of the theory via scattering Green's functions, derive the Green's functions for the layered case and calculate the impact of the cavity in terms of a local field correction in analogy to the known excess polarisabilities. Finally, we model the reflection at a continuous dielectric profile and illustrate the effect by applying the method to example profiles.

The Casimir force can be obtained from the electromagnetic stress tensor~\cite{Buhmann2013Dispersion} as a surface integral over the considered body. The Abraham stress tensor can be obtained in terms of dyadic Greens function~\cite{Friedrich}
\begin{eqnarray}
%\lefteqn{ 
{\bf{T}}({\bf{r}}) = -\frac{\hbar}{\pi} \int\limits_0^\infty \mathrm d \xi \left\lbrace \frac{\xi^2}{c^2}\varepsilon(\mi\xi) {\bf{G}}({\bf{r}},{\bf{r}},\mi\xi) +\frac{\tilde{{\bf{G}}}({\bf{r}},{\bf{r}},\mi\xi)}{\mu(\mi\xi)}%\right.}\nonumber \\
%&&\left.
-\frac{1}{2} \operatorname{tr}\left[ \frac{\xi^2}{c^2}\varepsilon(\mi\xi){\bf{G}}({\bf{r}},{\bf{r}},\mi\xi) +\frac{1}{\mu(\mi\xi)}\tilde{{\bf{G}}}({\bf{r}},{\bf{r}},\mi\xi)\right]\right\rbrace \, , \label{eq:tens}
\end{eqnarray}
where $ \tilde{{\bf{G}}}({\bf{r}},{\bf{r}},\omega)$ denotes the contributions from the magnetic field
\begin{equation}
 \tilde{{\bf{G}}}({\bf{r}},{\bf{r}},\omega) =\left. \nabla\times {\bf{G}}({\bf{r}},{\bf{r}}',\omega)\times \nabla'\right|_{{\bf{r}}'\to{\bf{r}}} \, .
\end{equation}
For infinite plates it is sufficient to introduce the force density per unit area
\begin{equation}
 {\bf{f}} = \frac{\mathrm d {\bf{F}}}{\mathrm d A} =  {\bf{e}}_z \cdot\left(\left. {\bf{T}}({\bf{r}})\right|_{z=b} -\left. {\bf{T}}({\bf{r}})\right|_{z=b'}\right) \, , \label{eq:CP}
\end{equation}
where $b$ and $b'=b-d_1$ denotes the positions of the right and left boundary of the considered plate, respectively. 
The dyadic Green function ${\bf{G}}({\bf{r}},{\bf{r}}',\omega)$ is the fundamental solution of the vector Helmholtz equation for purely dielectric systems \cite{Scheel2008}
\begin{equation}
 \left[\nabla\times\nabla\times - \frac{\omega^2}{c^2}\varepsilon({\bf{r}},\omega)\right]{\bf{G}}({\bf{r}},{\bf{r}}',\omega) = \boldsymbol{\delta}({\bf{r}}-{\bf{r}}') \, .
\end{equation}
\

A similar method can be applied to calculate the van der Waals interaction.  We consider two polarisable particles at positions ${\bf r}_1$ and ${\bf r}_2$ with polarisabilities $\alpha_1$ and $\alpha_2$. By applying fourth-order perturbation theory to the interaction Hamiltonian $H=-\hat{\bf{d}}_1 \cdot \hat{\bf{E}}({\bf r}_1)-\hat{\bf{d}}_2 \cdot \hat{\bf{E}}({\bf r}_2)$ the van der Waals potential can be found as \cite{Buhmann2013Dispersion}
\begin{eqnarray}
U_{vdW}({\bf r}_1, {\bf r}_2) = -\frac{\hbar \mu_0^2}{2\pi} \int\limits_0^\infty \mathrm d\xi\,\xi^4 \alpha_1(\mi\xi)\alpha_2(\mi\xi) \operatorname {tr} \left[{\bf G(r}_1, {\bf r}_2, \mi\xi)\cdot {\bf G(r}_2, {\bf r}_1, \mi\xi)\right] \, .
\label{eq:vdw}
\end{eqnarray}
By comparing Eqs.~(\ref{eq:tens}) and (\ref{eq:vdw}) one observes that Casimir force requires the coincidence limit (${\bf{r}}'\to{\bf{r}}$), whereas the van der Waals potential depends on the Green's tensor at two  positions. Thus, the Casimir force is governed by the reflections of virtual photons and the van der Waals force by their transmissions.

To address the influence of an inhomogeneous dielectric profile on the van der Waals force, we separate the profile into nine sections as illustrated in Fig.~\ref{fig:Csketch2}: five regions with spatially constant dielectric function and four with continuous profiles.

We start our investigations by neglecting the inhomogeneous profile regions. As mentioned before the important quantities are the transmission and reflections at these interfaces, which leads us to write the influence of the cavities in terms of reflection coefficients and to perform the transition to the cavities with continuous profiles via exchanging the hard-boundary refection coefficients with the ones determined for a continuous dielectric profile. 
The situation for the Casimir force is nearly the same. However, in this case we have to consider two additional layers denoting the plates representing the interacting objects and have to calculate the reflection coefficients at both sides of one of these plates, e.g. the left one illustrated in Fig.~\ref{fig:Csketch}.
\section{Green's functions for planar multi layered systems and local field corrections for planar cavities}
We assume the cavity around the particles to be small compared to the separation of the particles, $d_1,d_2\ll l$ and that the latter is small enough such that the interaction is nonretarded. This allows us to approximate the Green's function for the five-layer system in terms of a local-field corrected bulk Green's function. 

The Green's function of a system involving inhomogeneous media can be described by the bulk Green's function modified with the local field correction factors arising from the reflection and transmission of the electromagnetic field through the various layers of inhomogeneities of the system \cite{Buhmann2013Dispersion}. That is, for the five-layer system considered here, the full Green's tensor can be written as, similar to the local-field correction for cavities \cite{Sambale07}
\begin{eqnarray}
{\bf G(r, r'}, \omega) =\int\limits_0^\infty\mathrm d \kappa \,\mathrm e^{-\mi\kappa( z- z')}\left.\tilde{t}(\kappa)\right|_{{\bf{r}}} \cdot {\bf G}^{(1)}(z,z',\kappa, \omega)\cdot \left.\tilde{t}(\kappa)\right|_{{\bf{r}}'} \,, \label{eq:locfield}
\end{eqnarray}
where $\left.\tilde{t}\right|_{{\bf{r}}}$ and $\left.\tilde{t}\right|_{{\bf{r}}'}$ represent the generalized transmission coefficients close to the final and source point, respectively, while
${\bf G}^{(1)}{\bf (r, r'}, \omega)$ denotes the scattering Green's tensor for a bulk system. The transmission and reflection coefficients marked with a tilde denotes the complete ones including multiple reflections, whereas the ones without a tilde are the ordinary Fresnel coefficients at a single interface.

In order to estimate the complete van der Waals interaction, we write down the scattering Green's function for a single interface and obtain the multiple reflection coefficients for the multiple reflections. The scattering Green's tensor for a source situated in layer $1$ and field in layer $2$ is given by \cite{Buhmann2013Dispersion}
\begin{eqnarray}
{\bf G}{\bf (r, r'}, \mi\xi) = \frac{\mi}{8\pi^2} \int\limits_0^\infty \frac{\mathrm d^2k^{\parallel}}{k^\perp}\mathrm e^{\mi {\bf{k}}^\parallel \cdot ({\bf{r}}-{\bf{r}}')-\mi(k_2^\perp z-k_1^\perp z')} \sum_{\sigma=s,p} t_{12}^\sigma {\bf {e}}_2^{\sigma -}  {\bf {e}}_1^{\sigma -}\,, 
\end{eqnarray}
for $z < 0$ and $z' > 0$, where $k^\parallel$ and $k_j^\perp$ are the components of the wave vector parallel and perpendicular to the interface, respectively, ${\bf {e}}_j^{\sigma -}$ are the unit vectors for the polarisations given by
\begin{equation}
 {\bf {e}}_j^{s -} =\begin{pmatrix}
                          \sin\varphi \\-\cos\varphi \\0
                         \end{pmatrix}
\, ,\quad 
 {\bf {e}}_j^{p -} =\frac{1}{k_j}\begin{pmatrix}
                          k_j^\perp \cos\varphi \\k_j^\perp\sin\varphi \\k^\parallel
                         \end{pmatrix}\, .
\end{equation}
We shifted to imaginary frequencies $\omega=\mi\xi$. We set the transversal coordinates to zero, $x=x'=y=y'=0$. Thus, the $k$ integration can be transformed via polar coordinates $\mathrm d^2 k^\parallel = \sin\varphi \mathrm d \varphi \mathrm d k^\parallel$ and the angle integration can be performed leading
\begin{eqnarray}
%\lefteqn{
{\bf G} {\bf (r, r'}, \mi\xi) = \frac{1}{8\pi^2} \int\limits_0^\infty \frac{\mathrm d k^{\parallel}k^{\parallel}}{\kappa_1^\perp}\mathrm e^{-(\kappa_2^\perp z-\kappa_1^\perp z')}%}\nonumber \\ 
 %&&\times 
 \left[\pi t^s_{12} \begin{pmatrix}
   1    & 0                & 0   \\
   0               & 1   & 0  \\
   0               & 0                &0
\end{pmatrix} +t_{12}^p \frac{\pi c^2}{\xi^2\sqrt{\varepsilon_1\varepsilon_2}} \begin{pmatrix}
   {\kappa^\perp_1}{\kappa^\perp_2}    & 0                & 0   \\
   0               & {\kappa^\perp_1}{\kappa^\perp_2}   & 0  \\
   0               & 0                & -2{k^{\parallel}}^2 
   \end{pmatrix}  \right]\, ,
\end{eqnarray}
with the imaginary part of the perpendicular wave vectors in each layer $\kappa_j^2 = \varepsilon_j \xi^2/c^2+{k^\parallel}^2$. In the non-retarded limit, we have $k^\parallel \gg\xi/c $. Thus, the contribution of the scattering of $p$-polarised waves dominates the process
\begin{eqnarray}
{\bf G} {\bf (r, r'}, \mi\xi) = \frac{c^2}{8\pi\xi^2\sqrt{\varepsilon_1\varepsilon_2}} \int\limits_0^\infty \mathrm d \kappa\,\mathrm e^{-\kappa( z- z')} t_{12}^p \kappa^2 \operatorname{diag}\left(1,1,-2\right)  \, , \label{eq:twolayer}
\end{eqnarray}
where we have used $\kappa =\kappa_1^\perp=\kappa_2^\perp = k^\parallel$ meaning that $k^\parallel$ dominates the dispersion relation. 
Further, in the non-retarded limit, the transmission coefficient simplifies to the Fresnel coefficient \cite{Tomas1995,Chew,Buhmann2013Dispersion}
\begin{equation}
 t_{12}^p = \frac{2\varepsilon_1}{\varepsilon_1+\varepsilon_2} \, ,
\end{equation}
and consequently it becomes independent from the wave vector $k$. By further considering a homogeneous medium, which means that $\varepsilon_1=\varepsilon_2$ the transmission coefficient simplifies to $1$ and the integral can be performed~\cite{Buhmann2013Dispersion}
\begin{eqnarray}
{\bf G} {\bf (r, r'}, \mi\xi) = \frac{c^2}{4\pi\xi^2\varepsilon} \frac{1}{( z- z')^3} \operatorname{diag}\left(1,1,-2\right)  \, . \label{eq:bulk1}
\end{eqnarray}
The result can be compared with the non-retarded bulk Green's tensor~\cite{Buhmann2013Dispersion}
\begin{equation}
 {\bf{G}}({\bf{r}},{\bf{r}}',i\xi) = \frac{c^2}{4\pi\xi^2\varepsilon\varrho^3}\left[{\bf{I}}-3{\bf{e}}_{\varrho}{\bf{e}}_{\varrho}\right] \, , \label{eq:bulk2}
\end{equation}
with the three-dimensional unit matrix $\bf{I}$, the relative coordinate ${\boldsymbol{\varrho}}={\bf{r}}-{\bf{r}}'$, its absolute value $\varrho=\left|\boldsymbol{\varrho}\right|$ and its unit vector ${\bf{e}}_\varrho =\boldsymbol{\varrho}/\varrho$. By again setting $x=x'=y=y'=0$ in Eq.~(\ref{eq:bulk2}), we indeed recover Eq.~(\ref{eq:bulk1}).
Thus, the results are consistent.

Coming back to the original situation, we consider the three layer system. By assuming a large central layer we can neglect the multiple scattering inside this layer due to the large damping by the propagation through it. Thus, we can construct the complete transmission by the product of the transmission coefficient entering the middle layer $t_{12}$ and exiting it $t_{23}$, leading to identify the generalized transmission coefficients from Eq.~(\ref{eq:locfield}) by
\begin{equation}
 \left.\tilde{t}\right|_{{\bf{r}}} = t_{12} \, , \qquad \left.\tilde{t}\right|_{{\bf{r}}'}=t_{23} \, .
\end{equation}
Assuming that both particles are the same, also the shape of the cavities will equal and the transmission out of the middle layer can be transformed to
\begin{equation}
 t_{23} = 1+ r_{23} = 1-r_{32}=2-t_{32} \, .
\end{equation}
Thus, for a symmetric profile, we further simplify the generalized transmission coefficients to
\begin{equation}
 \left.\tilde{t}\right|_{{\bf{r}}} = t_{12} \, , \qquad \left.\tilde{t}\right|_{{\bf{r}}'}=2-t_{12} \, .
\end{equation}

In order to adapt the transmission through a two-layer system to the aimed five-layer scenario, the transmission coefficient has to be modified with respect to multiple reflections at the additional interfaces. Regarding the five-layer problem depicted in Fig.~\ref{fig:Csketch2} together with the assumption of small cavity sizes $d=2R_C\ll l$ the multiple reflection inside the layer separating both particles can be neglected due to the strong absorption along the propagation inside this region $\propto \mathrm e^{-\kappa_2 l}$. This leads to the restriction of the consideration of multiple reflection only inside the cavity with width $d$. For a single cavity, it is sufficient to consider a three-layer system, where a centred medium with dielectric function $\varepsilon_1$ and width $d$ is in contact with two infinite extended layers with dielectric functions $\varepsilon_{2\pm}$ (where ``+'' denotes the right medium and ``-'' the left medium). We consider a particle centred in the middle layer and estimate the transmission of an electromagnetic wave created at the particle into the right medium $\varepsilon_{2+}$. All optical paths starting inside the cavity and terminating in this medium can be written as
\begin{eqnarray}
\lefteqn{ \tilde{t}_{12+} = t_{12+}} \nonumber \\
&& +r_{12+} p r_{12-} p t_{12+} +r_{12+} p r_{12-} p r_{12+} p r_{12-} p t_{12+} + \dots\nonumber\\
&& + r_{12-} p t_{12+}+r_{12-} p r_{12+} p r_{12-} p t_{12+} +\dots \, ,
\end{eqnarray}
with the propagation along the cavity $p=\mathrm e^{-\kappa d}$. The first term denotes the part which is directly transmitted, the second line denotes all odd parts, which are initially sent towards right interface and the third line denotes the even parts, which starts with a reflection at the left interface before all are transmitted into the third layer. This equation can be written into a geometric series
\begin{eqnarray}
\tilde{t}_{12+} = \left[1+r_{12-} \mathrm e^{-\kappa d}\right] \left[\sum_{n=0}^\infty \left(r_{12-}r_{12+}\mathrm e^{-2\kappa d}\right)^n\right] t_{12+} = \frac{1+r_{12-} \mathrm e^{-\kappa d}}{1-r_{12-}r_{12+}\mathrm e^{-2\kappa d}}t_{12+} \, ,
\end{eqnarray}
again the first term collects even and odd paths, the second term is the sum over all multiple reflections and finally the transmission into the right layer. By considering the left and right medium to be equal $\varepsilon_{2-}=\varepsilon_{2+}$ and using the relation between reflection and transmission coefficients for a layered system $t=1+r$ the local-field corrections simplify to
\begin{equation}
 \tilde{t} = \frac{1+r}{1-r\mathrm e^{-\kappa d}} \, , \label{eq:lfvdw}
\end{equation}
with the Fresnel reflection coefficient $r=(\varepsilon-1)/(\varepsilon+1)$. In the limit of vanishing cavity size ($d\to 0$) the local-field correction can be further simplified to
\begin{equation}
 \tilde{t}=\frac{1+r}{1-r} \, .
\end{equation}
Thus it can be further simplified to $\tilde{t}=\varepsilon$. Summarising, the local-field correction, Eq.~(\ref{eq:locfield}), reads
\begin{eqnarray}
{\bf G(r, r'}, \omega) =\int\limits_0^\infty\mathrm d \kappa \,\mathrm e^{-\mi\kappa( z- z')}\frac{1+r}{1-r\mathrm e^{-\kappa d}}   {\bf G}^{(1)}(z,z',\kappa, \omega) \left(2-\frac{1+r}{1-r\mathrm e^{-\kappa d}} \right) \,, 
\end{eqnarray}
and can be further simplified to 
\begin{eqnarray}
{\bf G(r, r'}, \omega) =\varepsilon(2-\varepsilon) \int\limits_0^\infty\mathrm d \kappa \,\mathrm e^{-\mi\kappa( z- z')}{\bf G}^{(1)}(z,z',\kappa, \omega)\,, \nonumber\\
\end{eqnarray}
by assuming a vanishing cavity.

In order to describe the corresponding situation for the Casimir force, we again start with the scattering Green's function for a two layered system. In contrast to the van der Waals case, the source and final points have to be located in the same layer in this case. Hence, we begin our calculation with a similar expression for the Green's function and need to replace the transmission coefficients by the corresponding expression for reflections, leading to \cite{Buhmann2013Dispersion}
\begin{eqnarray}
{\bf G}{\bf (r, r'}, \omega) = \frac{\mi}{8\pi^2} \int\limits_0^\infty \frac{\mathrm d^2k^{\parallel}}{k^\perp}\mathrm e^{\mi {\bf{k}}^\parallel \cdot ({\bf{r}}-{\bf{r}}')-\mi(k_2^\perp z-k_1^\perp z')} r_{12}^p {\bf {e}}_1^{p +}  {\bf {e}}_1^{p -}\,, 
\end{eqnarray}
where in the non-retarded limit the reflection coefficient for $s$-polarised wave vanishes directly for a dielectric medium. 

\begin{figure}[htb]
\centering
\includegraphics[width=0.6\textwidth]{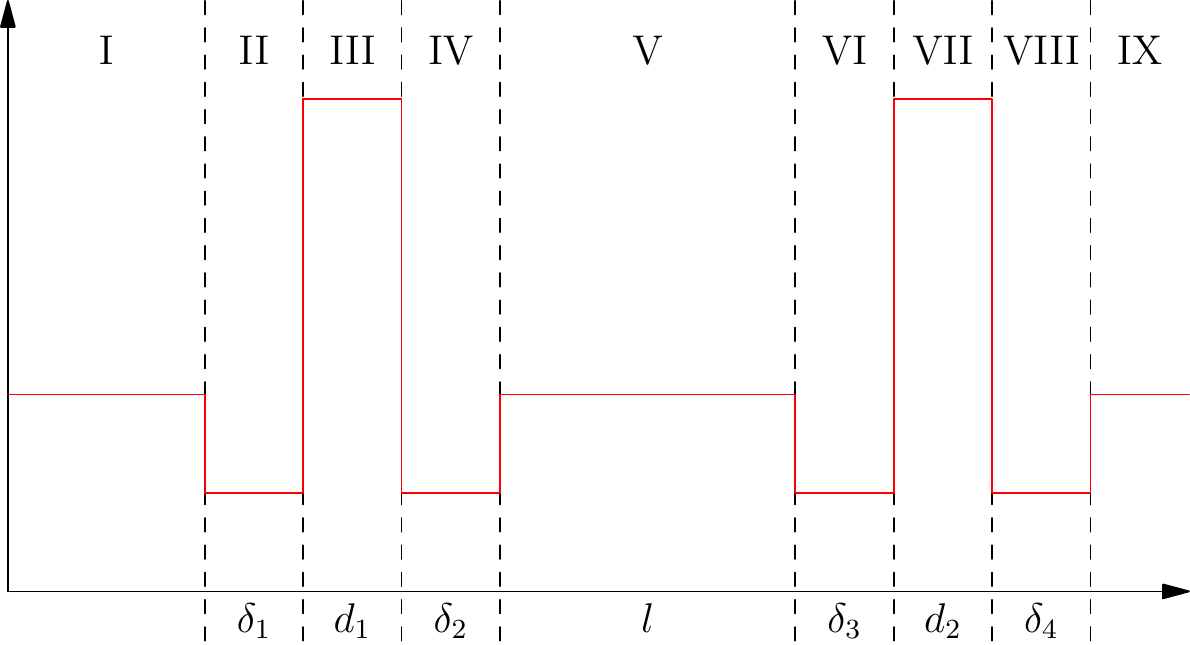}
 \caption{Sketch of the nine-layer system for the Casimir force.}
 \label{fig:ninelayer}
\end{figure}

In this case, we have to consider a nine-layer system depicted in Fig.~\ref{fig:ninelayer} and estimate the Casimir pressure acting on the left plate. In Fig.~\ref{fig:ninelayer} the two plates of thickness $d_1$ and $d_2$ are illustrated as layer III and VII. Each of them is located in a cavity marked by layer II and IV with thickness $\delta_1$ and $\delta_2$ for the left plate and layer VI and VIII with thickness $\delta_3$ and $\delta_4$. Layer I and IX have an infinite thickness. Both subsystems are separated by the middle layer V of thickness $l$.  Before we continue with the estimation of the local-field correction for this case, we recall the motivation for this analysis which is the estimation of the impact of a continuous dielectric profile for the one-dimensional analog for the real cavity models. This means that the two subsystems consisting of the layers I to IV and of the layers VI to IX are fixed. Thus, we can treat these two subsystems as two effective layers and can map the situation onto the well-known case of a planar cavity resulting the simple expression for the Casimir force~\cite{Buhmann2013Dispersion}
\begin{equation}
 {\bf{f}} = \frac{\hbar}{2\pi^2} \int\limits_0^\infty\mathrm d \xi \int\limits_0^\infty \mathrm d k^\parallel \kappa^\perp k^\parallel \frac{r_p^+r_p^-\mathrm e^{-2\kappa^\perp l}}{1-r_p^+r_p^-\mathrm e^{-2\kappa^\perp l}} {\bf{e}}_z \, , \label{eq:Casimir}
\end{equation}
where $r_p^\pm$ denotes the reflection coefficients for $p$-polarised waves at the right, which is the subsystem 2 consisting of the layers VI-IX, and left interface, which is subsystem 1 with layers I-IV, according to the three layer system discussed for the van der Waals force and we used that the reflection coefficients for $s$-polarised waves vanish.

Let us consider the local-field correction for the left subsystem (layer I to IV). In terms of the nine-layer system the effective reflection coefficient at the interface between the fifth and the fourth layer has to be determined. It will be constructed iteratively starting with effective reflection between the third and second interface~\cite{Chew}
\begin{equation}
 \tilde{r}_{32}= r_{32} +\frac{t_{32}r_{21}t_{23}\mathrm e^{-\kappa_{\mathrm{II}} \delta_1}}{1-r_{32}r_{23}\mathrm e^{-2\kappa_{\mathrm{II}}\delta_1}} \, ,
\end{equation}
where $\kappa_{\mathrm{II}}$ denote the imaginary part of the wave vector in the second layer. Starting from this effective reflection coefficient, one can continue with the next interface
\begin{equation}
 \tilde{r}_{43} = \frac{r_{43} +\tilde{r}_{32} \mathrm e^{-2\kappa_{\mathrm{III}}d_1}}{1-r_{43} \tilde{r}_{32} \mathrm e^{-2\kappa_{\mathrm{III}}d_1}} \, ,
\end{equation}
and finally
\begin{equation}
 \tilde{r}_{54} = \frac{r_{54} +\tilde{r}_{43} \mathrm e^{-2\kappa_{\mathrm{IV}}\delta_2}}{1-r_{54} \tilde{r}_{43} \mathrm e^{-2\kappa_{\mathrm{IV}}\delta_2}} \, ,
\end{equation}
with the wave vectors in the third and fourth layer $\kappa_{\mathrm{III}}$ and $\kappa_{\mathrm{IV}}$. Assuming that $\varepsilon_{\mathrm{II}}=\varepsilon_{\mathrm{IV}}=1$ and $\varepsilon_{\mathrm{I}}=\varepsilon_{\mathrm{V}}=\varepsilon$ which leads to equal wave vectors in the corresponding layers and equal distances for the cavity layers ($\kappa_{\mathrm{II}}=\kappa_{\mathrm{IV}}=\kappa$ and $\delta_1=\delta_2=R_C$) and thus only two elementary reflection coefficients are important, the one between the cavity and the slab $r_{23}=r_{43}=r_1$ and between the cavity and the medium $r_{21}=r_{45}=r$ and the generalized reflection coefficient simplifies to
\begin{eqnarray}
 \tilde{r} &=& -\frac{1+r_1^2 \mathrm e^{-2\kappa R_C}}{r_1^4\mathrm e^{-4\kappa R_C} +r r_1^3 \mathrm e^{-3\kappa R_C} -r r_1\mathrm e^{-3\kappa R_C} +2 r_1^2\mathrm e^{-2\kappa R_C}+1} \left[  r^2r_1( r_1^2-1) \mathrm e^{-2 \kappa_{\mathrm{III}} d_1-\kappa  R_C}\right.\nonumber\\
 &&\left.+(r r_1^4+r_1) \mathrm e^{-2 \kappa_{\mathrm{III}} d_1-2 \kappa R_C}+(r r_1^2-r) \mathrm e^{-2 \kappa_{\mathrm{III}} d_1-3 \kappa R_C}+r_1^3\mathrm e^{-2 \kappa_{\mathrm{III}} d_1-4 \kappa R_C}\right.\nonumber\\
 &&\left.+(r r_1^2-r_1) \mathrm e^{-2 \kappa R_C}+r_1^2 r\mathrm e^{-2 \kappa_{\mathrm{III}}d_1} -r_1^3\mathrm e^{-4 \kappa R_C} +r\right]\nonumber\\
 &&\times \left[r_1^4 \mathrm e^{-2\kappa_{\mathrm{III}}d_1-2\kappa R_C} + r_1^3 r \mathrm e^{-2\kappa_{\mathrm{III}}d_1 -\kappa R_C}+r_1^2\mathrm e^{-2\kappa R_C} -r_1 r\mathrm e^{-2\kappa_{\mathrm{III}}d_1-\kappa R_C}+r_1^2\mathrm e^{-2\kappa_{\mathrm{III}}d_1}+1\right]^{-1} \, .\label{eq:refl}
\end{eqnarray} 
The generalized reflection coefficient at the right interface can be determined analogously leading the same result by exchanging the reflection coefficient at the slab to the other materials $r_1=(\varepsilon_1-1)/(\varepsilon_1+2)\mapsto r_2 =(\varepsilon_2-1)/(\varepsilon_2+1)$. Now, we restrict ourselves to the case that both slabs consist of equal materials leading to the same reflection coefficient.

\section{Reflection at an inhomogeneous boundary}

With respect to an arbitrary one-dimensionally spatial susceptibility profile, $\varepsilon(z,\omega)$, the resulting reflection coefficient can be obtained by solving the Riccati differential equation~\cite{Chew}
\begin{eqnarray}
 R'(z) = -2\kappa \sqrt{\varepsilon(z,\omega) +\left(\frac{k_\parallel}{\kappa}\right)^2} R(z)-\frac{1}{4} \frac{\varepsilon'(z,\omega)}{\varepsilon(z,\omega)}\frac{2k_\parallel^2+\kappa^2\varepsilon(z,\omega)}{k_\parallel^2+\kappa^2\varepsilon(z,\omega)}\left[1-R^2(z)\right] \, ,
\end{eqnarray}
for $p$-polarised waves, with $\kappa=\omega/c$. We omit the discussion of $s$-polarised waves, because they vanish in the nonretarded limit. This equation has a unique solution with the initial condition $R(z\to-\infty) = 0$ and result in the right-sided reflection coefficients $R^+(z)$ at the position $z$. The capital letter $R$ denotes the reflection coefficient at the position $z$. In contrast the small letter $r$ denotes the reflection coefficient at the specific distance which is required for the local-field corrections. Analogously, we find the left sided reflection coefficients by using the  relation 
\begin{equation}
R^-(z) = -R^+(z) \, .
\end{equation}
\begin{figure}[t!]
\centering 
\includegraphics[width=0.4\columnwidth]{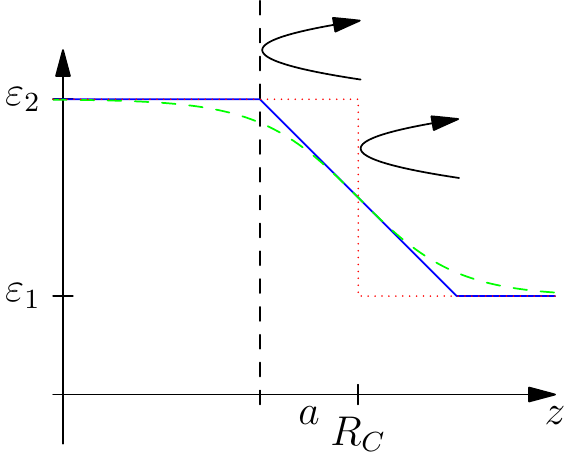}
 \caption{Sketch of the different spatial dependent dielectric functions for the hard boundary (red dotted curve), the linear profile (blue curve) and the non-linear profile (green dashed curve).The black arrows illustrate the considered reflections which take place for the hard boundary at the cavity radius $R_C$ and for the functional profiles at their ends. Shifted by a distance $a$ compared to the hard-boundary case.}\label{fig:scenarios}
\end{figure}

Figure~\ref{fig:scenarios} illustrates the different profiles of inhomogeneity, where we want to describe the reflection process. Due to the spatial dependence of the profile the point of reflection changes by the distance $a$ which we assume to be equal for all types of profiles and define its value by the crossing point of a linear profile reaching the final value as depicted in Fig.~\ref{fig:scenarios}.

Again, we apply the non-retarded limit to estimate the reflection coefficients. Then the Riccati differential equation simplifies to
\begin{equation}
  R'(z) = -2 k_\parallel R(z) -\frac{1}{2} \frac{\varepsilon'(z,\omega)}{\varepsilon(z,\omega)} \left[1-R^2(z)\right]  \, .\label{eq:ric}
\end{equation}
 This equation can be solved analytically for vanishing wave vector ($k_\parallel =0$) and results in
\begin{equation}
 R(z) = \tanh \left\lbrace -\frac{1}{2} \left[\ln\varepsilon(z,\omega) - \ln\varepsilon(-a,\omega)\right]\right\rbrace \, , \label{eq:sol}
\end{equation}
which can be simplified further to
\begin{equation}
 R(z_{end}) = \frac{\varepsilon_2-\varepsilon_1}{\varepsilon_2+\varepsilon_1} \, ,\label{eq:refl0}
\end{equation}
by assuming that the profile connects both dielectric functions as depicted in Fig.~\ref{fig:scenarios}. This result denotes the ordinary Fresnel reflection coefficient in agreement to the considered case of non-propagating waves. Due to this fact, we can figure out the impact of the linear term describing the absorption of light by the propagation through the finite profile of the length $2a$.

Corresponding to the profile plotted in Fig.~\ref{fig:scenarios}, we choose two different spatial profiles for the dielectric function: \\(i) a linear function
\begin{eqnarray}
 \varepsilon(z,\omega) = \varepsilon_1(\omega)+\left[\varepsilon_2(\omega)-\varepsilon_1(\omega) \right]\times \begin{cases}
                          1 & \text{for}\  z<-a\\
                          \frac{a-z}{2a} & \text{for} \left|z\right| \le a\\
                          0 & \text{else}
                         \end{cases} \, ,
\end{eqnarray}
and (ii) a Thomas-Fermi distributed profile
\begin{equation}
 \varepsilon(z,\omega) = \varepsilon_1(\omega)+\left[\varepsilon_2(\omega)-\varepsilon_1(\omega) \right]\left(1+\mathrm e^{\frac{2z}{a}}\right)^{-1} \, , \label{eq:profTF}
\end{equation}
approximating typical profiles, see Ref.~\cite{Walter14}. Both profile have the same slope in the origin ($z=0$). The width of the inhomogeneous profiles are $2a$. Thus, the profiles end at the point $a$. The boundary conditions are for case (i) $R(-a)=0$ and for case (ii) $R(z\to-\infty)=0$. 

\begin{figure}[htb]
\centering
 \includegraphics[width=0.6\textwidth]{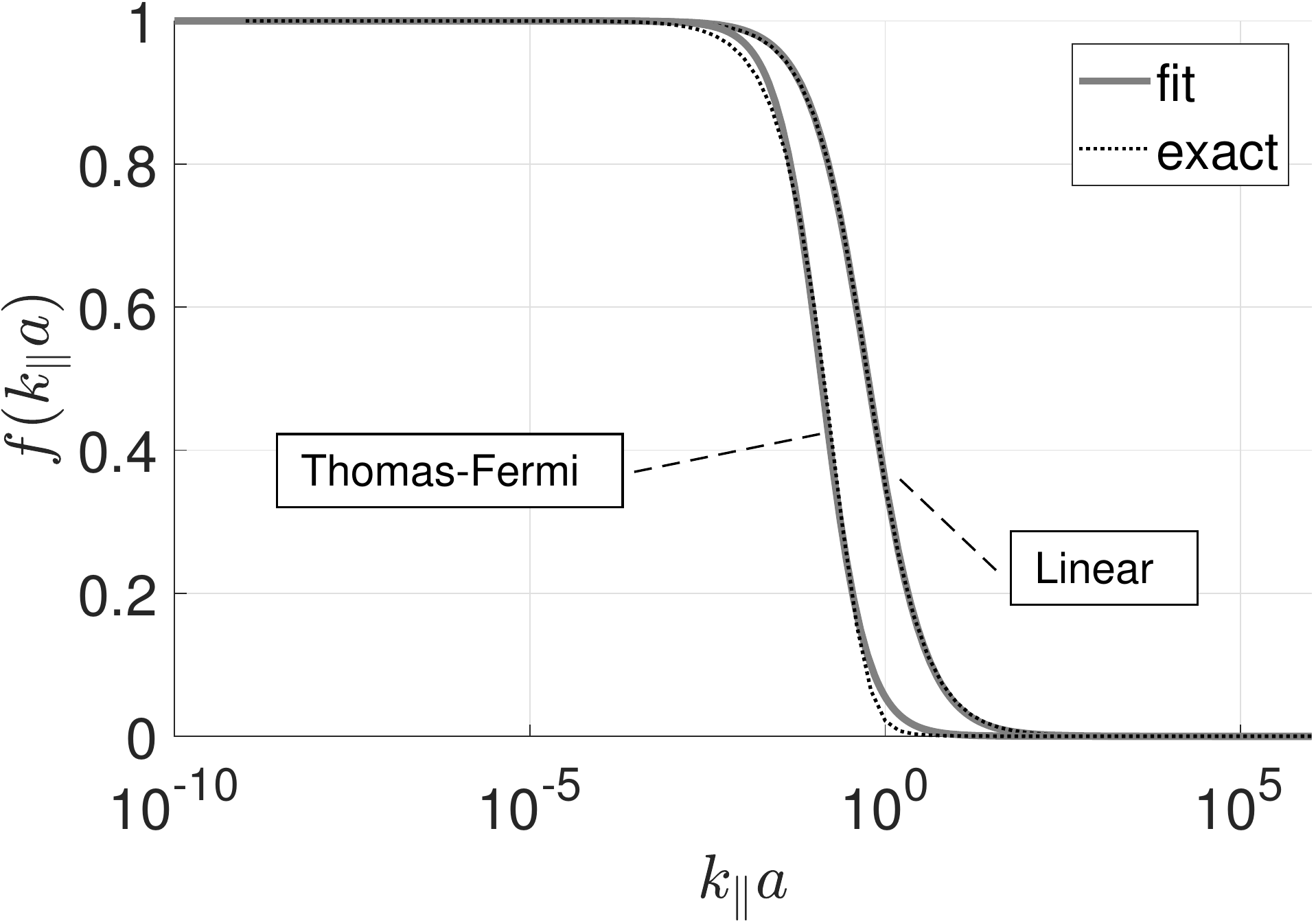}
 \caption{Geometric dependence of the reflection coefficients for both profiles - Thomas-Fermi and Linear - with the exact solution of the Riccati (black dotted line) equation and the corresponding approximations (grey solid lines).}\label{fig:refl}
\end{figure}

Numerical solutions of the Riccati differential equation~(\ref{eq:ric}) are depicted in Fig.~\ref{fig:refl} where we evaluated the reflection coefficient at the end of inhomogeneous profile which is for the linear profile the distance $a$ and for the Thomas-Fermi distribution we increased the final point to $3a$ reaching the final value with a deviation of less than 1\%. It can be observed that the $k_\parallel$ dependence is related to the thickness of the inhomogeneous region $2a$. Several checks with different parameters showed that the resulting curves only depend on the product of the wave vector and the corresponding length scale of the inhomogeneity. The depicted results can be understood by introducing the wavelength $\lambda=2\pi/k_\parallel$ leading to the ratio of the wavelength and the inhomogeneity's thickness as the relevant quantity. It can be seen that if this ratio is small $a/\lambda<1$ the specific profile does not matter to the result and the solution behaves like a hard boundary. In the other case where the ratio is large $a/\lambda>1$ the reflection decreases to zero. The interesting region is denoted by the case when the wavelength is comparable to the thickness $\lambda\approx a$. From the numerical simulations one can conclude that the final reflection coefficient separates into a product of two terms: one describes the dielectric properties and the other the geometric properties
\begin{equation}
 r\left[k,\varepsilon_1(\omega),\varepsilon_2(\omega), a\right] = \frac{\varepsilon_2-\varepsilon_1}{\varepsilon_2+\varepsilon_1} \cdot  f(ka) \, ,\label{eq:model}
\end{equation}
satisfying the $k_\parallel\to0$ limit, Eq.~(\ref{eq:refl0}). For both investigated cases the expression of the result are the same with different parameters. In analogy to the solution of the Riccati equation, Eq.~(\ref{eq:sol}) and in agreement with the numerical results depicted in Fig.~\ref{fig:refl}, we approximate the $k_\parallel$-dependence of the reflection coefficient by
\begin{equation}
 f(ka) \approx \frac{1}{2}\left[1-\tanh\left(\frac{\ln(ka)-\lambda_1}{\lambda_2}\right)\right] \, . \label{eq:approx}
\end{equation}
 The resulting parameters are given in Table~\ref{tbl:fit}, which prefectly match the curves with an accuracy of $\approx 100$\% for the linear profile and of 99.96\% for the Thomas-Fermi distribution. Note that the reflection coefficients for a linear profile have also been determined in Refs.~\cite{Prachi2018,PhysRevA.96.032123,PhysRevD.84.065028} based on methods similar to ours.

\begin{table}[htb]
\centering
 \begin{tabular}{l|cc}
  Profile & $\lambda_1$ & $\lambda_2$  \\\hline
Linear & $-0.555$ & $2.028$  \\
Thomas-Fermi & $-2.067$&  $1.452$ 
 \end{tabular}
\caption{Fitting parameter for both profiles (linear and Thomas-Fermi distributed) based on the approximation equation~(\ref{eq:approx}).% and the thickness of the inhomogeneous layer for helium in water resulting from the fit, Eq.~(\ref{eq:fitprof}).
}\label{tbl:fit}
\end{table}

\section{Results and discussion}

\begin{figure}[tb]
\centering
 \includegraphics[width=0.4\columnwidth]{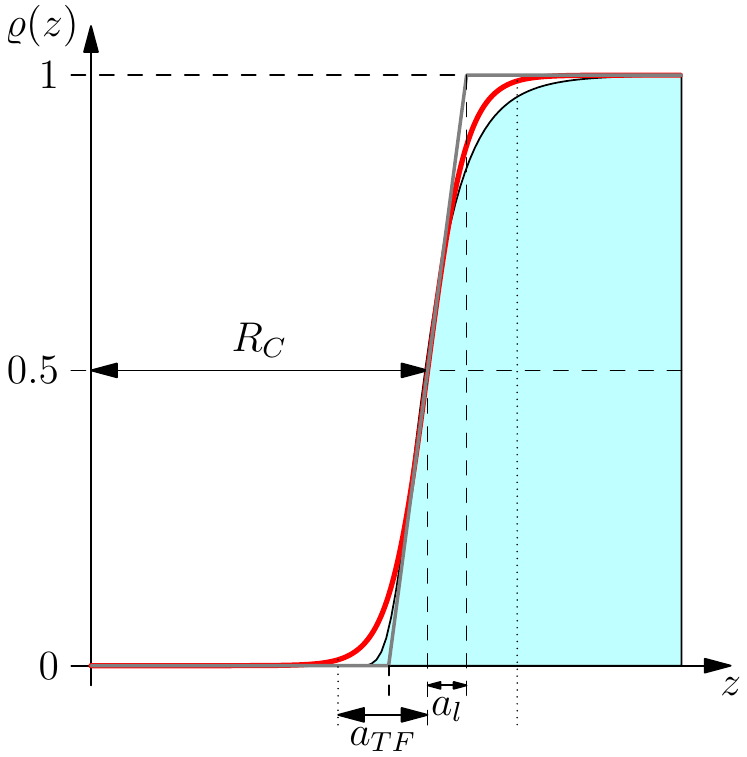}
 \caption{ 
Spatial density profile of water surrounding a helium atom with the cavity radius $R_C$. Simulated profile enclosed the blue area which is approximated by a Thomas-Fermi distribution (red line) with the corresponding thickness of the inhomogeneous region $2a_{TF}$ and by a linear profile (grey line) with the thickness $2a_l$.}\label{fig:cavity}
\end{figure}

Applying the local-field correction, Eq.~(\ref{eq:locfield}), to the van der Waals potential~(\ref{eq:vdw}) one finds
\begin{eqnarray}
\lefteqn{U_{vdW}(z, z') =}\nonumber\\
&&-\frac{3\hbar}{64\pi^3\varepsilon_0^2} \int\limits_0^\infty \mathrm d\xi\frac{\alpha_1(i\xi)\alpha_2(i\xi)}{\varepsilon_1(i\xi)\varepsilon_2(i\xi)} \int\limits_0^\infty \mathrm d \kappa\mathrm d \kappa'\,\kappa^2\mathrm e^{-\kappa( z- z')} \tilde t(\kappa)[2-\tilde t(\kappa)] \mathrm e^{-\kappa'( z- z')} \tilde t(\kappa')[2-\tilde t(\kappa')] {\kappa'}^2 \, ,
\label{eq:vdw2}
\end{eqnarray}
with the local-field corrected transmission coefficients~(\ref{eq:lfvdw}), and the reflection coefficient $r$ has to be evaluated for the inhomogeneous profile which is modelled by Eqs.~(\ref{eq:model}) and (\ref{eq:approx}).

In order to illustrate the theory we consider helium atoms solved in water. Helium's polarisability were taken from Ref.~\cite{DEREVIANKO2010323} and the dielectric function for water from Ref.~\cite{Elbaum273Kwater}. 
Assuming the permittivity to be proportional
to the distribution of water around the helium atom \cite{Walter14},
the resulting dielectric profile is depicted in Fig.~\ref{fig:cavity}. 
The simulated profile is fitted to a Thomas-Fermi distribution
\begin{equation}
 \varrho(z) = \left(1+\mathrm e^{-\alpha(z-R_C)}\right)^{-1} \, , \label{eq:fitprof}
\end{equation}
with the cavity radius $R_C=1.71$\AA$\,$  and the profile's slope $\alpha=10.1$\AA$^{-1}$, via a least square algorithm. A comparison of this function with the profiles used to solve the Riccati differential equation~(\ref{eq:profTF}), relates the fitting parameter $\alpha$ to the generalised profile width $a_l=2/\alpha$ directly denoting the thickness for the linear profile. Using the same value to determine the thickness of the Thomas-Fermi distributed profile would result an error which can be improved by increasing the layer size. To do so, we define the thickness in this case by a threshold of 99\% [$\varrho(a_{TF}) = 0.99$] leading to $a_{TF} =- \ln(0.0101)/\alpha$. The resulting parameters are 0.198\AA\, for the linear profile and 0.455\AA\, for the Thomas-Fermi profile.

\begin{figure}[tb]
\centering
 \includegraphics[width=0.6\columnwidth]{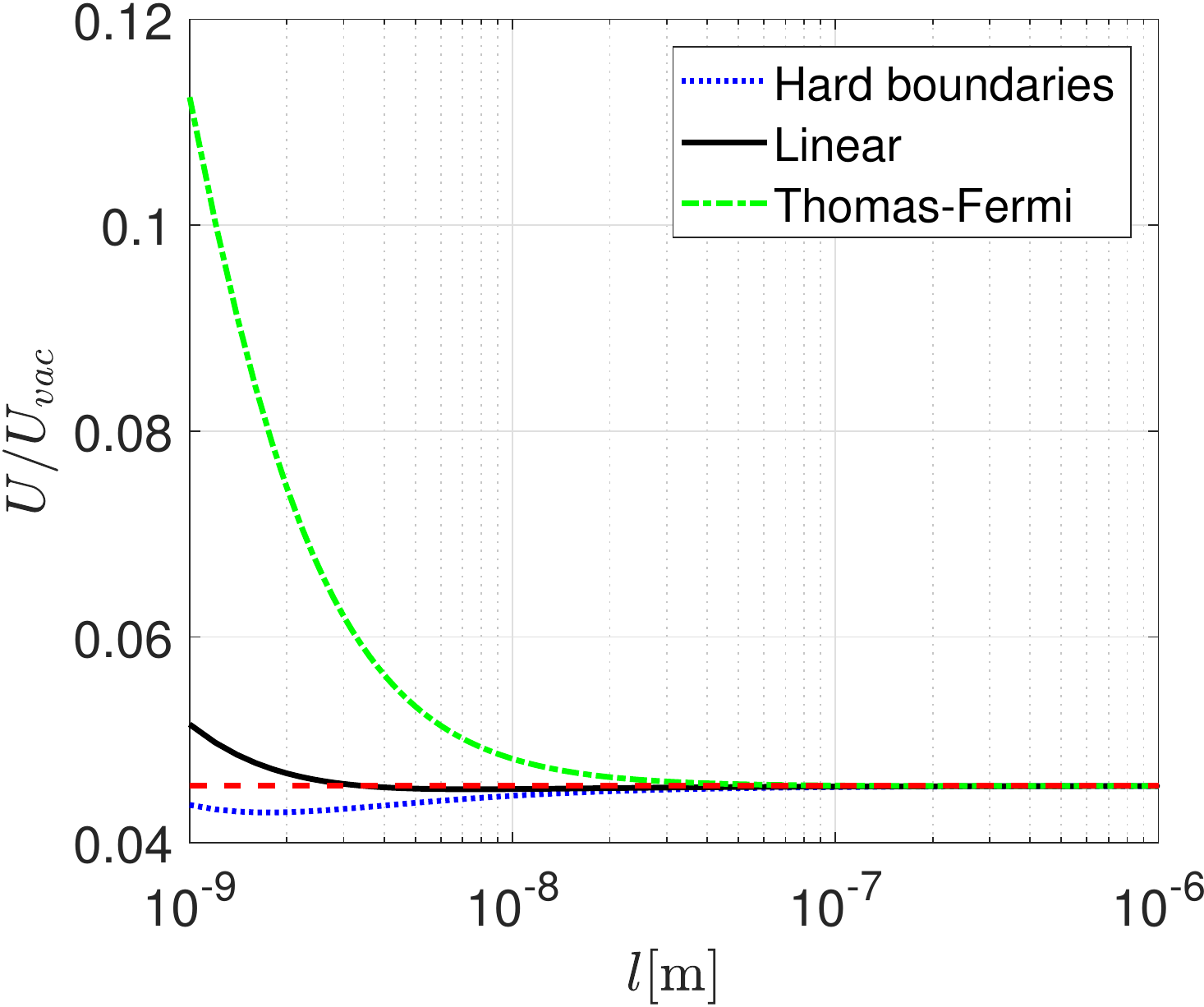}
 \caption{Relative van der Waals potentials for the cavity model with hard boundaries (blue dotted line), linear boundaries (black solid line) and Thomas-Fermi distributed boundary profiles (green dashed-dotted line) with respect to the vacuum potential. The general damping of the potential caused by the absorption of water (factor $\approx20$) can be observed (red dashed line).}\label{fig:vdW}
\end{figure}

The van der Waals potentials for helium atoms in water are depicted in Fig.~\ref{fig:vdW}. A planarly layered cavity profile surrounding two helium atoms is considered, which has an equivalent dimension as the spherical cavity. This means that the atoms are 1.71\AA \,behind the water-vacuum interfaces. In the figure, the relative potential with respect to the vacuum potential over the thickness of the intermediate layer $l$ is plotted. It can be observed that for large separations ($30 \,\rm{nm}$, which is approximately ten-times the extension of the cavity layer), the interaction becomes independent of the microscopic details of the cavity profile and agrees with the predicted screened and local-field corrected result~(\ref{eq:vdw2}). For shorter distances the hard-boundary profile (blue dotted line) results in a further reduction whereas the potentials for the more realistic linear (solid black line) and for the Thomas-Fermi distributed profiles (green dashed dotted line) increase for shorter distances, in the latter case by a factor of up to two. The hierarchy of different results can be understood from the fact that the van der Waals potential is due to photons exchanged between the two atoms. The transmission of these photons is least effective for the hard-boundary case due to the strong reflectivity and most effective for the smooth and more weakly reflecting Thomas-Fermi profile.

For the Casimir force, we apply the method to a similar example, where we replace the helium atoms by two helium plates of finite thickness 
set to the van der Waals radius of helium 
$d_1=d_2=d=1.43$\AA~\cite{doi:10.1021/ic501364h}, similar values can be found in Refs.~\cite{YZ1995,doi:10.1021/jp2094438}.
 The dielectric function of these plates can be estimated via the Clausius--Mossotti relation~\cite{jackson1998classical,Johannes}
\begin{equation}
 \varepsilon= \frac{1+2\alpha/(4\pi\varepsilon_0 d^3)}{1-\alpha/(4\pi\varepsilon_0 d^3)} \, ,
\end{equation}
entering the reflection coefficient
\begin{equation}
 r_1=\frac{\varepsilon-1}{\varepsilon+1} \, .
\end{equation}
\begin{figure}[htb]
\centering
 \includegraphics[width=0.6\columnwidth]{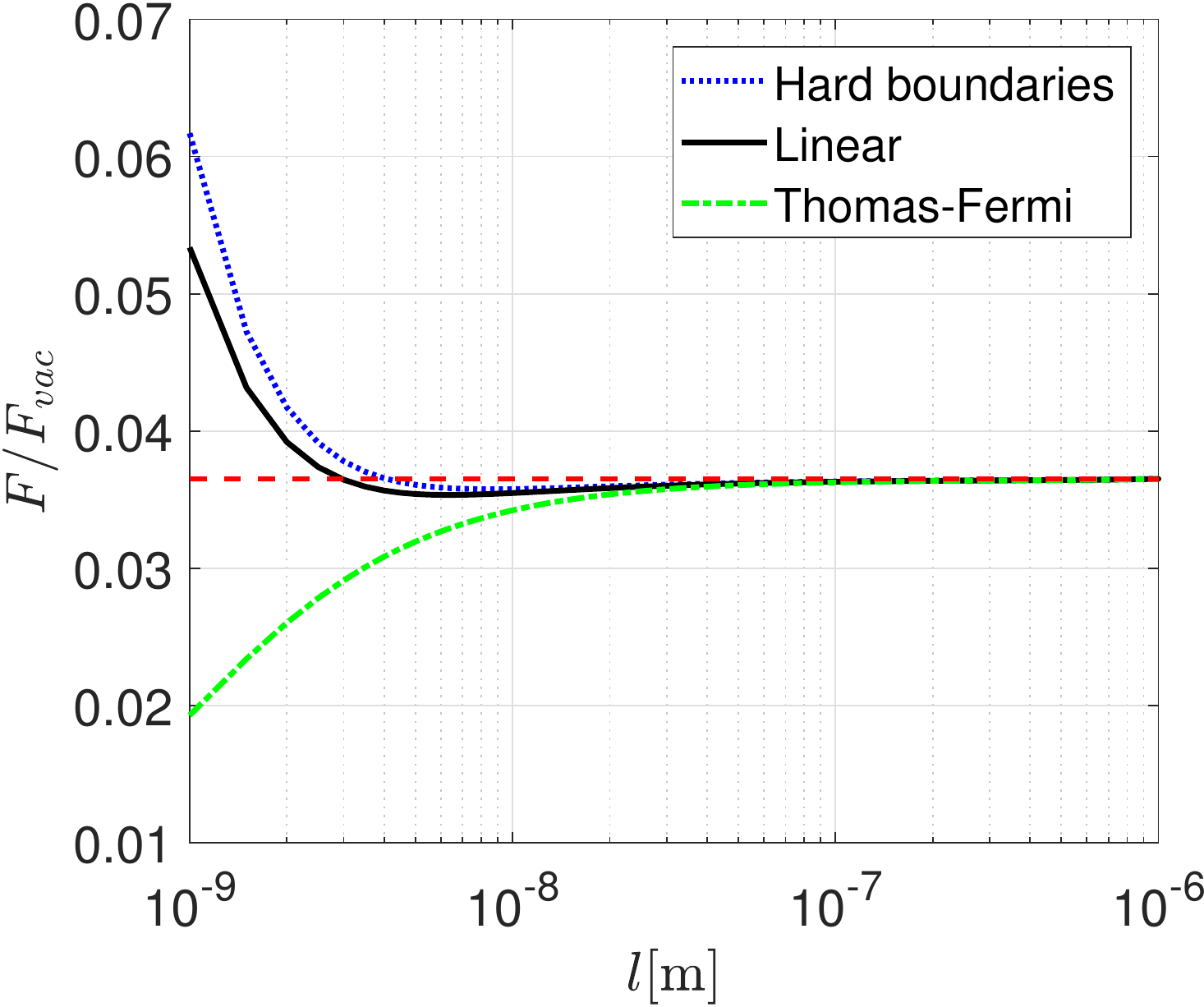}
 \caption{Relative impact of the cavity boundaries on the Casimir force for an helium plate embedded in water relative to the force in vacuum. Without any cavity corrections (red dashed line), with hard boundaries (blue dotted line), with the linear profile (black solid line) and with the Thomas-Fermi distributed profile (green dashed-dotted line).}\label{fig:C}
\end{figure}

By approximating the reflection coefficients for the cavities via Eq.~(\ref{eq:approx}) and applying the result to the local-field corrected Casimir force, Eq.~(\ref{eq:Casimir}) together with Eq.~(\ref{eq:refl}), the Casimir force acting on two helium plates in water can be obtained. The results are depicted in Fig.~\ref{fig:C}. 
Again, the local-field corrected screening due to the water can be observed in the long-range ($> 30\,\rm{nm}$) limit of all models, where all models coincide. Analogously to the van der Waals potential, the models strongly differ on shorter
scales. Surprisingly, the hard boundary model (blue dotted line) results in the strongest force. The linear profile (black solid line) yields a slight increase of the force similar to the van der Waals case. However, the Thomas-Fermi distributed boundary profile (green dashed dotted line) yields a force which is further reduced in comparison to the long-range result. The differences in the results can be explained from the fact that the Casimir force arises from a reflection of photons, which is strongest for the hard-boundary case and weakest for the Thomas-Fermi profile.

\section{Conclusions}
The aim of this article is the illustration of the impact of a cavity with a continuous boundary profile on dispersion interactions in media. To this end, we investigated an equivalent situation with a one-dimensional cavity and derived the local-field correction induced by the cavity simplifying the theory to be applicable for known two and three layer cases. Further, we approximated the Fresnel reflection coefficient at a continuous profile with a fitted function which is valid for the cases of a linear profile or a Thomas-Fermi distributed profile. To illustrate the impact on the relevant dispersion forces we applied the method to the case of a two helium atoms in water (van der Waals potential) and for two helium plates embedded in water (Casimir force). We found that the influence is relevant at small distances. On larger scales the impact of the continuous profile vanishes and the hard-boundary case is a suitable description. To account for more realistic three-dimensional systems, an adaptation of this theory to three-dimensional spherical objects is required, which will be subject of further investigations.

\section*{Acknowledgments}
We gratefully acknowledge support by the German Research Council (grant
BU1803/6-1, S.Y.B. and J.F.) the Research Innovation Fund by the University of Freiburg
(S.Y.B., J.F. and M.W.) and the Freiburg Institute for Advanced Studies (S.Y.B.). 
We acknowledge financial support from the Research Council of Norway (Project 250346 and 243642) and from COST Action MP1406 'MultiscaleSolar' (Project 39969).

\bibliographystyle{unsrt}  
\bibliography{Bib}  %%% Remove comment to use the external .bib file (using bibtex).

\begin{thebibliography}{10}

\bibitem{McLachlan387}
A.~D. McLachlan.
\newblock Retarded dispersion forces between molecules.
\newblock {\em Proceedings of the Royal Society of London A: Mathematical,
  Physical and Engineering Sciences}, 271(1346):387--401, 1963.

\bibitem{Casimir48}
H.~B.~G. Casimir and D.~Polder.
\newblock The influence of retardation on the london-van der waals forces.
\newblock {\em Phys. Rev.}, 73:360--372, Feb 1948.

\bibitem{Casimir482}
H.~B.~G. Casimir.
\newblock On the attraction between two perfectly conducting plates.
\newblock {\em Proc. K. Ned. Akad. Wet.}, 51:793, 1948.

\bibitem{Grisenti1999}
R.~E. Grisenti, W.~Sch\"ollkopf, J.~P. Toennies, G.~C. Hegerfeldt, and
  T.~K\"ohler.
\newblock Determination of {A}tom-{S}urface {V}an der {W}aals {P}otentials from
  {T}ransmission-{G}rating {D}iffraction {I}ntensities.
\newblock {\em Phys. Rev. Lett.}, 83:1755--1758, 1999.

\bibitem{Arndt1999}
M.~Arndt, O.~Nairz, J.~Vos-Andreae, C.~Keller, G.~van~der Zouw, and
  A.~Zeilinger.
\newblock Wave-particle duality of {C}60 molecules.
\newblock {\em Nature}, 401(6754):680--682, 1999.

\bibitem{Juffmann2012}
T.~Juffmann, A.~Milic, M.~M\"ullneritsch, P.~Asenbaum, A.~Tsukernik,
  J.~T\"uxen, M.~Mayor, O.~Cheshnovsky, and M.~Arndt.
\newblock Real-time single-molecule imaging of quantum interference.
\newblock {\em Nature Nanotechn.}, 7:297 -- 300, 2012.

\bibitem{Brand2015}
C.~Brand, J.~Fiedler, T.~Juffmann, M.~Sclafani, C.~Knobloch, S.~Scheel,
  Y.~Lilach, O.~Cheshnovsky, and M.~Arndt.
\newblock A {G}reen's function approach to modeling molecular diffraction in
  the limit of ultra-thin gratings.
\newblock {\em Ann. Phys. (Berlin)}, 527:580--591, 2015.

\bibitem{pitaevskii}
I~E Dzyaloshinskii, E~M Lifshitz, and Lev~P Pitaevskii.
\newblock General theory of van der waals' forces.
\newblock {\em Soviet Physics Uspekhi}, 4(2):153, 1961.

\bibitem{Scheel2008}
S.~Scheel and S.~Buhmann.
\newblock Macroscopic quantum electrodynamics - concepts and applications.
\newblock {\em Act. Phys. Slov.}, 58(5), 2008.

\bibitem{Buhmann2013Dispersion}
S.Y. Buhmann.
\newblock {\em Dispersion Forces I: Macroscopic Quantum Electrodynamics and
  Ground-State Casimir, Casimir--Polder and van der Waals Forces}.
\newblock Springer Tracts in Modern Physics. Springer Berlin Heidelberg, 2013.

\bibitem{Gecko}
Kellar Autumn, Yiching~A. Liang, S.~Tonia Hsieh, Wolfgang Zesch, Wai~Pang Chan,
  Thomas~W. Kenny, Ronald Fearing, and Robert~J. Full.
\newblock Adhesive force of a single gecko foot-hair.
\newblock {\em Nature}, 405:681--685, 2000.

\bibitem{Friedrich}
Friedrich~Anton Burger, Johannes Fiedler, and Stefan~Yoshi Buhmann.
\newblock Zero-point electromagnetic stress tensor for studying casimir forces
  on colloidal particles in media.
\newblock {\em EPL (Europhysics Letters)}, 121(2):24004, 2018.

\bibitem{messiah2014quantum}
A.~Messiah.
\newblock {\em Quantum Mechanics}.
\newblock Dover Books on Physics. Dover Publications, 2014.

\bibitem{Johannes}
Johannes Fiedler, Priyadarshini Thiyam, Anurag Kurumbail, Friedrich~A. Burger,
  Michael Walter, Clas Persson, Iver Brevik, Drew~F. Parsons, Mathias
  Bostr{\"o}m, and Stefan~Y. Buhmann.
\newblock Effective polarizability models.
\newblock {\em The Journal of Physical Chemistry A}, 121(51):9742--9751, 2017.
\newblock PMID: 29185741.

\bibitem{Walter14}
Alexander Held and Michael Walter.
\newblock {Simplified continuum solvent model with a smooth cavity based on
  volumetric data}.
\newblock {\em The Journal of Chemical Physics}, 141(17):174108, 2014.

\bibitem{Henkel2008}
C.~Henkel, G.~Boedecker, and M.~Wilkens.
\newblock Local fields in a soft matter bubble.
\newblock {\em Applied Physics B}, 93(1):217--221, Oct 2008.

\bibitem{Sambale07}
Agnes Sambale, Stefan~Yoshi Buhmann, Dirk-Gunnar Welsch, and Marin-Slodoban
  Toma\ifmmode~\check{s}\else \v{s}\fi{}.
\newblock Local-field correction to one- and two-atom van der waals
  interactions.
\newblock {\em Phys. Rev. A}, 75:042109, Apr 2007.

\bibitem{Tomas1995}
M.~S. Toma\ifmmode~\check{s}\else \v{s}\fi{}.
\newblock Green function for multilayers: Light scattering in planar cavities.
\newblock {\em Phys. Rev. A}, 51:2545--2559, Mar 1995.

\bibitem{Chew}
W.C. Chew.
\newblock {\em Waves and Fields in Inhomogeneous Media}.
\newblock Institute of Electrical \& Electronics Engineers (IEEE), 1995.

\bibitem{Prachi2018}
Prachi Parashar, Kimball~A. Milton, Yang Li, Hannah Day, Xin Guo, Stephen~A.
  Fulling, and Ines Cavero-Pelaez.
\newblock Quantum electromagnetic stress tensor in an inhomogeneous medium.
\newblock {\em arXiv:1804.04045}, 2018.

\bibitem{PhysRevA.96.032123}
Itay Griniasty and Ulf Leonhardt.
\newblock Casimir stress inside planar materials.
\newblock {\em Phys. Rev. A}, 96:032123, Sep 2017.

\bibitem{PhysRevD.84.065028}
Kimball~A. Milton.
\newblock Hard and soft walls.
\newblock {\em Phys. Rev. D}, 84:065028, Sep 2011.

\bibitem{DEREVIANKO2010323}
Andrei Derevianko, Sergey~G. Porsev, and James~F. Babb.
\newblock Electric dipole polarizabilities at imaginary frequencies for
  hydrogen, the alkali–metal, alkaline–earth, and noble gas atoms.
\newblock {\em Atomic Data and Nuclear Data Tables}, 96(3):323 -- 331, 2010.

\bibitem{Elbaum273Kwater}
Michael Elbaum and M.~Schick.
\newblock Application of the theory of dispersion forces to the surface melting
  of ice.
\newblock {\em Phys. Rev. Lett.}, 66:1713--1716, Apr 1991.

\bibitem{doi:10.1021/ic501364h}
J\"urgen Vogt and Santiago Alvarez.
\newblock van der waals radii of noble gases.
\newblock {\em Inorganic Chemistry}, 53(17):9260--9266, 2014.
\newblock PMID: 25144450.

\bibitem{YZ1995}
Youxue Zhang and Zhengjiu Xu.
\newblock Atomic radii of noble gas elements in condensed phases.
\newblock {\em American Mineralogist}, 80(7-8):670--675, 1995.

\bibitem{doi:10.1021/jp2094438}
Uwe Hohm and Ajit~J. Thakkar.
\newblock New relationships connecting the dipole polarizability, radius, and
  second ionization potential for atoms.
\newblock {\em The Journal of Physical Chemistry A}, 116(1):697--703, 2012.
\newblock PMID: 22098409.

\bibitem{jackson1998classical}
J.D. Jackson.
\newblock {\em Classical Electrodynamics}.
\newblock Wiley, 1998.

\end{thebibliography}


\end{document}